\documentclass[aps,prb,twocolumn,superscriptaddress]{revtex4-1}
\usepackage{longtable}
\usepackage{morefloats}
\usepackage[dvips]{graphicx}
\usepackage{color}
\usepackage{epsfig,graphicx,amsfonts,amsbsy}
\usepackage{amsmath,amsfonts,amsthm,amssymb}
\usepackage{appendix}
\usepackage{makeidx}
\usepackage{url}
\usepackage[bookmarksnumbered,pdfpagelabels=true,plainpages=false,colorlinks=true,linkcolor=blue,citecolor=red,urlcolor=blue]{hyperref}

\begin{document}
\title{Symmetry and localization properties of defect modes in magnonic superlattices}
\author{R. A. Gallardo}
\affiliation{Departamento de F\'{i}sica, Universidad T\'{e}cnica Federico Santa Mar\'{i}a, Avenida Espa\~{n}a 1680, Valpara\'{i}so, Chile}
\affiliation{Center for the Development of Nanoscience and Nanotechnology (CEDENNA), 917-0124 Santiago, Chile}
\author{T. Schneider}
\affiliation{Helmholtz-Zentrum Dresden -- Rossendorf, Institut of Ion Beam Physics and Materials Research, Bautzner Landstr. 400, 01328 Dresden, Germany}
\affiliation{Department of Physics, Technische Universit{\"a}t Chemnitz, Reichenhainer Str. 70, 09126 Chemnitz, Germany}
\author{A. Rold\'an-Molina}
\affiliation{Universidad de Ays\'en, Ovispo Vielmo 62, Coyhaique, Chile}
\author{M. Langer}
\affiliation{Helmholtz-Zentrum Dresden -- Rossendorf, Institut of Ion Beam Physics and Materials Research, Bautzner Landstr. 400, 01328 Dresden, Germany}
\affiliation{Paul Scherrer Institut, 5232 Villigen PSI, Switzerland}
\author{A. S. N\'{u}\~{n}ez}
\affiliation{Departamento de F\'{i}sica, Facultad de Ciencias F\'{i}sicas y Matem\'{a}ticas, Universidad de Chile, Casilla 487-3, 8370415 Santiago, Chile}
\affiliation{Center for the Development of Nanoscience and Nanotechnology (CEDENNA), 917-0124 Santiago, Chile}
\author{K. Lenz}
\affiliation{Helmholtz-Zentrum Dresden -- Rossendorf, Institut of Ion Beam Physics and Materials Research, Bautzner Landstr. 400, 01328 Dresden, Germany}
\author{J. Lindner}
\affiliation{Helmholtz-Zentrum Dresden -- Rossendorf, Institut of Ion Beam Physics and Materials Research, Bautzner Landstr. 400, 01328 Dresden, Germany}
\author{P. Landeros}
\affiliation{Departamento de F\'{i}sica, Universidad T\'{e}cnica Federico Santa Mar\'{i}a, Avenida Espa\~{n}a 1680, Valpara\'{i}so, Chile}
\affiliation{Center for the Development of Nanoscience and Nanotechnology (CEDENNA), 917-0124 Santiago, Chile}
\date{\today }
\pacs{}
\keywords{magnonic crystals, magnonic supercells, spin waves, plane wave method, flat-bands}

\begin{abstract}
Symmetry and localization properties of defect modes of a one-dimensional bi-component magnonic superlattice are theoretically studied.
The magnonic superlattice can be seen as a periodic array of nanostripes, where stripes with different width, termed as defect stripes, are periodically introduced.
By controlling the geometry of the defect stripes, a transition from dispersive to practically flat spin-wave defect modes can be observed inside the magnonic band gaps.
It is shown that the spin-wave profile of the defect modes can be either symmetric or antisymmetric, depending on the geometry of the defect. 
Due to the localized character of the defect modes, a particular magnonic superlattice is proposed, wherein the excitation of either symmetric or antisymmetric flat magnonic modes is enabled at the same time.
Also, it is demonstrated that the relative frequency position of the antisymmetric mode inside the band gap does not significantly change with the application of an external field, while the symmetric modes move to the edges of the frequency band gaps.
The results are complemented by numerical simulations, where an excellent agreement is observed between both methods.
The proposed theory allows exploring different ways to control the dynamic properties of the defect modes in metamaterial magnonic superlattices, which can be useful for applications on multifunctional microwave devices operating over a broad frequency range.
\end{abstract}

\maketitle

\section{Introduction}

The dynamic properties of spin waves (SWs) in magnonic devices with artificial periodic modulation of the magnetic or geometrical parameters have been a glowing research area in the last years.\cite{Vasseur96,Neusser08,Tacchi11,Tacchi12,Krawczyk08,Wang09,Wang10,Chumak12,Yu13,Krawczyk14}
The magnetic metameterials termed magnonic crystals (MCs) have been widely studied since their excitation spectrum presents magnonic band gaps (BGs), which can be controlled by external magnetic fields.\cite{Serga10,Kruglyak10,Lenk11}
These systems can also be created by an artificial modulation of the magnetic properties,\cite{Kruglyak06,Barsukov11,Obry13,Gallardo14}  or by the modification of the film geometry.\cite{Chumak09,Lee09,Ding11,Landeros12,Ciubotaru12,Klos12,Krawczyk13,Korner13,Gubbiotti14,Langer17,Gallardo18} 
The magnonic BGs are strongly dependent on the geometrical parameters of the periodic lattice, whose spatial range usually lies in the hundreds of nanometers.
Around these gaps, SWs can be excited in well defined allowed frequency bands, where, depending on the wave vector, the waves may have a standing or a propagating character.
In particular, the standing SWs occur at the borders of the Brillouin zones and therefore, such waves can only be excited at some specific wave vectors.
This characteristic makes it difficult to channel or guide the spin waves along specific regions, which is the key for applications in magnonic waveguides\cite{Lenk11,Krawczyk14} and tunable narrow passband SW filters.\cite{Di14a,Zhang16}

In the field of photonic crystals, it is well known that the incorporation of a local defect breaks the translational symmetry and electromagnetic modes can appear within the forbidden band gaps.\cite{Yablonovitch91,Joannopoulos97,Bayindir00,Braun06}
For instance, the addition of extra dielectric material in one of the unit cells gives rise to modes within the BGs that behave like a donor atom in a semiconductor, while the removal of dielectric material from the crystal produces acceptor-like modes.\cite{Yablonovitch91}
This extra degree of freedom allows manipulating and controlling the properties of light in dielectric metamaterials. 
Indeed, the defect-induced phenomena in photonic crystals have been applied for controlling the spontaneous light emission,\cite{Ogawa04,Englund05,Kao16} and trapping optical pulses.\cite{Amir14}
In analogy to photonic crystals, it has been also demonstrated that the controlled introduction of periodic defects in magnonic crystals induces defect modes inside the BGs that can beneficially enrich the SW band structure of the magnetic metamaterial.\cite{Di14a,Zhang16}
Here, the periodic lattice induces translational symmetry to the SWs that can be broken by a controlled introduction of periodic defects.
That is a change in the periodic structure redefines the unit cell in the same way as in a crystal with a complex unit cell. 
This system can be seen as a \emph{magnonic superlattice} (MSL), which consists of a periodic array of magnonic supercells.  

The emergence of defect modes being located within the band gap has been predicted\cite{Nikitov01,Kuchko05,Kruglyak06b} and recently observed experimentally in magnonic superlattices.\cite{Di14a,Zhang16} 
Multilayered ferromagnetic structures with variations in the magnetization, uniaxial anisotropy and/or thicknesses have been theoretically studied in backward volume (BV) geometry.\cite{Nikitov01,Kuchko05,Kruglyak06b}.
A theoretical analysis of short-wavelength perturbations in two-dimensional MCs with point defects was performed by Yang, et al.,\cite{Yan12a,Yan12b,Yang14,Xing15} where different configurations of the point defects were investigated. 
Defect-induced phenomena in one dimensional bi-component MCs with structural defects were more recently investigated by Brillouin light scattering (BLS) measurements and via numerical simulations,\cite{Di14a,Zhang16}  
where arrays of 250 nm wide Py stripes were fabricated in such a way that every tenth wire there is a defect wire having a larger width ranging from 300 to 500 nm. 
Since the recent experiments were performed in the Damon-Eshbach (DE) geometry at small wave vectors (around 10 $\mu$m$^{-1}$), the dynamic dipolar contribution inevitably must be taken into account.
Nevertheless, there is no theoretical description that considers the dynamic magneto-dipolar fields and hence, the current measurements have been compared only with micromagnetic and finite-element simulations.\cite{Chi14,Di14a,Zhang16}
Furthermore, the analysis of an arbitrary angle of the magnetization with respect to the symmetry axes of the crystal has not been addressed so far.
Besides, the evolution of the defect modes as a function of the external field has not been deeply explored, either.
These aspects clearly hinder a complete study of the dynamic properties on MSL structures so far, since there is no model available that considers a general way to introduce arbitrary arrays of periodic defects on the MC.

In this paper, the symmetry and localization properties of defect modes within one-dimensional bi-component magnonic superlattices are theoretically addressed and complemented with micromagnetic simulations. 
It is shown that by controlling the lattice parameter of the defect stripes, a transition from slight to almost null dispersion of the defect modes is observed. 
Besides, by changing the width of the defect stripes the nature of the symmetry  as well as the frequency of the defect modes can be modified. 
It is also demonstrated that the external field can change the relative position of the symmetric modes with respect to the BG, while the antisymmetric ones remain at the same relative frequency position.
The possibility of exciting both symmetric and antisymmetric defect modes at the same time is also proposed, which allows for observing the defect modes in straightforward way with ferromagnetic resonance (FMR) measurements.

\begin{figure}[t]
\includegraphics[width=0.45\textwidth]{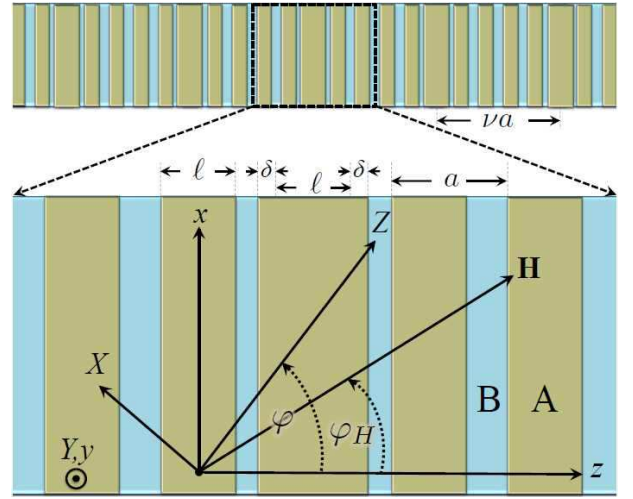}
\caption{Geometry of the one-dimensional bi-component magnonic superlattice composed of ferromagnetic materials A and B.
The lattice parameter of the periodic array of nanostripes is $a$, while $\nu a$ corresponds to the lattice parameter of the defect stripes, where $\nu$ represents the number of lattice repetitions that are necessary to form the magnonic superlattice. 
The width of the nanostripes (defects) is $\ell$ ($\ell +2\delta$).
The spin waves are assumed to propagate in $z$-direction, while the equilibrium magnetization (external field) form an angle $\varphi$ ($\varphi_{\rm h}$) with the $z$-axis.
The zoom denotes the unit cell of the MSL structure.}
\label{Fig1}
\end{figure}

\section{Theoretical description}

\begin{figure*}[t]
\includegraphics[width=\textwidth]{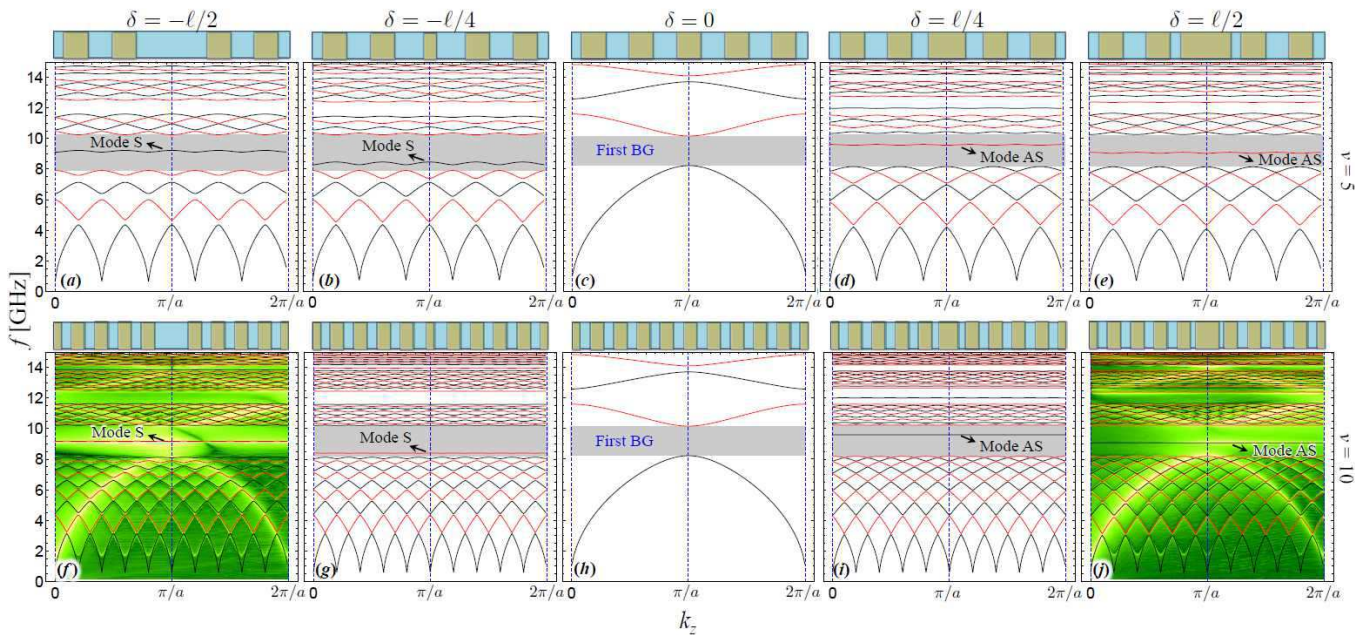} 
\caption{In ($a$)--($e$) the dispersion of a superlattice as given by $\nu=5$ is shown, while in ($f$)--($j$) the case $\nu=10$ is depicted. 
The parameter $\delta$ has been varied from $-\ell/2$ up to $\ell/2$ in such a way that the width ($\ell +2\delta$) of the defect ranges from $0$ to $a$. 
The illustration above each plot schematically shows the unit cell of the magnonic superlattice structure, whereas the gray zone depicts the first band gap.
In ($f$) and ($j$) the color code represents the numerical simulations, where the brighter color indicates a maximum of the response.
This response is given in log scale and corresponds to $\delta m = \sqrt{\delta m_x^2+\delta m_y^2+\delta m_z^2}$, where $\delta $ refers to the subtraction of the ground state before the fast Fourier transform.}
\label{Fig2}
\end{figure*}

By combining a defect-free lattice with a periodic array of stripes with different width, a one-dimensional bi-component magnonic superlattice is formed, as shown in Fig.\ \ref{Fig1}.
Here, the lattice parameter of the defect-free crystal is $a$, while the lattice parameter of the defect stripes is $\nu a$. 
Here, $\nu=1,2,3...$, is introduced to locate a defective wire at each $\nu$ repetition, allowing for a general description of the MSL.

The dynamics of the magnetic system is described by the Landau-Lifshitz (LL) equation $\mathbf{\dot{M}}(\mathbf{r};t)=-\gamma \mathbf{M}(\mathbf{r};t)\times\mathbf{H}^{\rm{e}}(\mathbf{r};t)$, where $\gamma$ is the absolute value of the gyromagnetic ratio, $\mathbf{M}(\mathbf{r};t)$ is the magnetization and $\mathbf{H}^{\rm{e}}(\mathbf{r};t)$ is the effective field.
For small magnetization deviations around the equilibrium state, both magnetization and effective field can be written as $\mathbf{M}(\mathbf{r};t)=M_{\rm{s}}(\mathbf{r})\hat{Z}+\mathbf{m}(\mathbf{r};t)$ and $\mathbf{H}^{\rm{e}}(\mathbf{r};t)=\mathbf{H}^{\rm{e0}}(\mathbf{r})+\mathbf{h}^{\rm{e}}(\mathbf{r};t)$, respectively.
Here, $M_{\rm{s}}(\mathbf{r})$ is the saturation magnetization, $\hat{Z}$ represents the equilibrium orientation of the magnetization and $\mathbf{m}(\mathbf{r};t)=m_{{X}}(\mathbf{r};t)\hat{X}+m_{Y}(\mathbf{r};t)\hat{Y}$ corresponds to the dynamic magnetization.
Besides, $\mathbf{H}^{\rm{e0}}(\mathbf{r})$ is the static part of the effective field and $\mathbf{h}^{\rm{e}}(\mathbf{r};t)$ is the time-dependent part.
Now, assuming a harmonic time dependence, $\mathbf{m}(\mathbf{r};t)=\mathbf{m}(\mathbf{r})e^{i\omega t}$, and neglecting the second order terms in $\mathbf{m}(\mathbf{r})$, the LL  equation can be written as
\begin{equation}
i(\omega/\gamma) m_{X}(\mathbf{r})=-m_{Y}(\mathbf{r})H^{\rm{e0}}_{Z}(\mathbf{r}) +M_{\rm{s}}(\mathbf{r})h^{\rm{e}}_Y(\mathbf{r})
\label{EQ1}
\end{equation}
and
\begin{equation}
i(\omega/\gamma) m_{Y}(\mathbf{r})=m_{X}(\mathbf{r})H^{\rm{e0}}_{Z}(\mathbf{r}) -M_{\rm{s}}(\mathbf{r})h^{\rm{e}}_X(\mathbf{r});
\label{EQ2}
\end{equation}
with $\omega$ being the angular frequency. 
Note that in Eqs.\ (\ref{EQ1}) and (\ref{EQ2}), the equilibrium conditions $M_{\rm{s}}(\mathbf{r})H^{\rm{e0}}_Y(\mathbf{r})=0$ and $M_{\rm{s}}(\mathbf{r})H^{\rm{e0}}_X(\mathbf{r})=0$ have been considered.
Now, the effective field is given  by $\mathbf{H}^{\rm{e}}(\mathbf{r})=\mathbf{H}+\mathbf{H}^{\rm{ex}}(\mathbf{r})+\mathbf{H}^{\rm{d}}(\mathbf{r})$, where $\mathbf{H}$ is the external field, $\mathbf{H}^{\rm{ex}}(\mathbf{r})$ is the exchange field and $\mathbf{H}^{\rm{d}}(\mathbf{r})$ is the dipolar field.
These fields are detailed in appendix \ref{ApA}.
According to Bloch's theorem, the dynamic magnetization components are expanded into Fourier series as $\mathbf{m}(\mathbf{r})=\sum_{\mathbf{G}}\mathbf{m}(\mathbf{G})e^{i\mathbf{(G+k)}\cdot\mathbf{r}}$, where $\mathbf{G}= G_{n}^{\nu} \hat{z}$ denotes the reciprocal lattice vector.
Here $G_{n}^{\nu}=(2\pi/\nu a) n$, where $n$, and $\nu$ are integer numbers.
The saturation magnetization and exchange length are respectively given by $M_{\rm{s}}(\mathbf{r})=\sum_{\mathbf{G}}M_{\rm{s}}(\mathbf{G})e^{i\mathbf{G}\cdot\mathbf{r}}$ and $\lambda_{\rm{ex}}(\mathbf{r})=\sum_{\mathbf{G}}\lambda_{\rm{ex}}(\mathbf{G})e^{i\mathbf{G}\cdot\mathbf{r}}$.
Here it is assumed that the leading material contrasts are associated to the saturation magnetization and exchange stiffness. 
Nevertheless, a contrast in anisotropies or the intrinsic damping can also modify the band structure of the spin waves as well as their relaxation time.
The detailed analysis of these cases is beyond the scope of this paper and in the following the study will be limited to variations in $M_{\rm{s}}(\mathbf{r})$ and $\lambda_{\rm{ex}}(\mathbf{r})$, which provides information capable of reproducing the measured band structure,\cite{Di14a,Zhang16} but not the lifetime of the modes.

Now, by including the effective fields in Eqs.\ (\ref{EQ1}) and (\ref{EQ2}), the LL equation can be converted into the following eigenvalue problem:\cite{Krawczyk08,Vasseur96}
\begin{equation}
\tilde{\mathbf{A}}\ \mathbf{m}_{\mathbf{G}}=i(\omega/\gamma) \ \mathbf{m}_{\mathbf{G}}
\label{BT}
\end{equation}
where $\mathbf{m}^{\rm T}_{\mathbf{G}}=[m_X(G_1),...,m_X(G_N),m_Y(G_1),...,m_Y(G_N)]$ is the eigenvector and $\tilde{\mathbf{A}}$ is given by  
\begin{eqnarray}
\tilde{\mathbf{A}}=\left( 
\begin{array}{cc}
 \tilde{\mathbf{A}}^{XX} 
& \tilde{\mathbf{A}}^{XY} 
\\
 \tilde{\mathbf{A}}^{YX} 
& \tilde{\mathbf{A}}^{YY} 
\end{array}\right).
\label{subM}
\end{eqnarray}
By using standard numerical methods and a convergence test to check the reliability of the results, the eigenvalues and eigenvectors of Eq.\ (\ref{BT}) can be obtained.
The matrix elements are given in appendix \ref{ApA}.


\section{Micromagnetic simulations}
Micromagnetic simulations were performed with the GPU-accelerated open-source code MuMax$^{3}$.\cite{Vansteenkiste14} 
The bi-component magnonic crystal was modeled as a 20 nm $\times$ 30 nm $\times$ 100 $\mu$m stripe. 
Periodic boundary conditions were applied to regain the thin film nature of the system. 
The stripe was discretized into 4 $\times$ 1 $\times$ 16384 cells, which results in a cell size of $5$ nm $\times$ $30$ nm $\times$ $6.1$ nm. 
The material parameters in the simulation were chosen as mentioned in Sec. \ref{Sec:results}. 
In addition, the Gilbert damping value of $0.01$ was chosen. 
Two kinds of simulations were performed for the magnonic supercell. 
At first the SW dispersion relation was calculated by applying a sinc-pulse in time and space.\cite{Venkat13}
In addition to the approach in Ref.\ \onlinecite{Venkat13}, the sinc-pulse was shifted in space by 30.5~nm with respect to the unit cell to also excite the totally antisymmetric SW modes.  
The resulting SW dispersion relations were obtained by performing a 2D Fast-Fourier-Transform for every lines of cells in the $z$-direction. 
Furthermore, the FMR response of the system has been simulated. 
Therefore, the time evolution of the system excited by a sinc-pulse in the time domain was recorded.\cite{McMichael05} 
To excite the antisymmetric SW modes as well, an additional linear offset was added. 
The SW frequencies were extracted as the summation of the spatial FFT in the time domain within of each cell. 

\section{Results and discussion}
\label{Sec:results}

To study the dynamic properties of the system, standard values of Cobalt and Permalloy are employed.\cite{Di14a}
Namely, the magnetic properties of material A resemble Permalloy (Ni$_{80}$Fe$_{20}$) which are: $M^{{\rm{A}}}_{\rm{s}}=730$ kA/m and $A_{\rm{ex}}^{{\rm{A}}}=1.1\cdot10^{-11}$ J/m.
On the other side, the magnetic properties of the material B correspond to Cobalt, i.e., $M^{{\rm{B}}}_{\rm{s}}=1100$ kA/m and $A_{\rm{ex}}^{{\rm{B}}}=2.5\cdot10^{-11}$ J/m.
Here, $A_{\rm{ex}}$ is the exchange constant and hence $\lambda_{\rm{ex}}=\sqrt{2A_{\rm{ex}}/4\pi M^2_{\rm{s}}}$ is the exchange length.
For both materials, an effective gyromagnetic ratio of $\gamma=0.0185556$ GHz/G and thickness $d=30$ nm are used.
Also, the lattice parameter of the defect-free crystal is $a=500$ nm and its width is $\ell=250$ nm.
At 200 reciprocal lattice vectors, a convergence of the numerical solutions of Eq.\ (\ref{BT}) is reached.

\begin{figure}[t]
\includegraphics[width=0.48\textwidth]{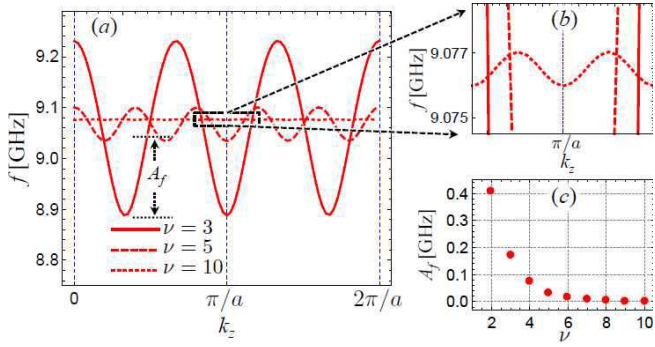}
\caption{Figure ($a$) illustrates the antisymmetric defect mode evaluated at $\delta= \ell/2$ for $\nu=3$, 5 and 10.
In ($b$) a zoom of the dispersion around of one boundary of the first Brillouin zone is shown.
Figure ($c$) depicts the oscillation amplitude $A_f$ as a function of $\nu$.}
\label{Fig4}
\end{figure}

Fig.\ \ref{Fig2} shows the SW dispersion of a MSL as created by $\nu=5$ [($a$)--($e$)] and $\nu=10$ [($f$)--($j$)] in the Damon-Eshbach geometry at $H=0$. 
Here, the equilibrium magnetization is given by $\varphi=\pi/2$ and the SW propagation is along the $z$-axis.
The parameter $\delta$ has been varied from $-\ell/2$ up to $\ell/2$ in such a way that the width ($\ell +2\delta$) ranges from $0$ to $a$.
In both cases, $\nu=5$ and $\nu=10$, a practically flat defect mode labeled as antisymmetric (AS) moves from the high frequency region into the first BG when $\delta>0$. 
As $\delta$ increases, this mode moves into the band gap and becomes localized around the center of the gap at $\delta=\ell/2$.
Conversely, if $\delta<0$, the symmetric (S) mode at the low-frequency edge of the first BG enters into the magnonic BG and becomes localized close to the center of the gap at $\delta=-\ell/2$.
Once both modes, S and AS, are located inside the BG they are characterized by a nearly flat dispersion. 
Overall, one can see that at higher values of $\nu$ the dispersion of the modes becomes flatter.  
Note that the case shown in Fig.\ \ref{Fig2}($j$) coincides with the system measured in Ref.\ \onlinecite{Di14a}.
Indeed, all parameters used in this paper are the same. 
Therefore, by comparing Fig.\ 3($b$) of Ref.\ \onlinecite{Di14a} with Fig.\ \ref{Fig2}($j$), one obtains an excellent agreement between them.

Figs.\ \ref{Fig2}($f$) and ($j$) show a comparison between the micromagnetic simulations and the theoretical results.
Overall, an excellent agreement is reached between both methods, which corroborates the validity of the theoretical model.
In the case $\nu=10$ depicted in Figs.\ \ref{Fig2}($f$)--($j$), it is clear that some defect modes also appear in the second BG. 
Nevertheless the behavior of these modes does not have a clear dependence with the geometrical parameter of the modified stripe.
For instance, at $\delta>0$ they are localized within the second band gap, while at $\delta<0$ the defect modes are localized around the second band-gap edges.
In what follows, the results are focused only on the defect modes localized within the first band gap.

\begin{figure}[t]
\includegraphics[width=0.48\textwidth]{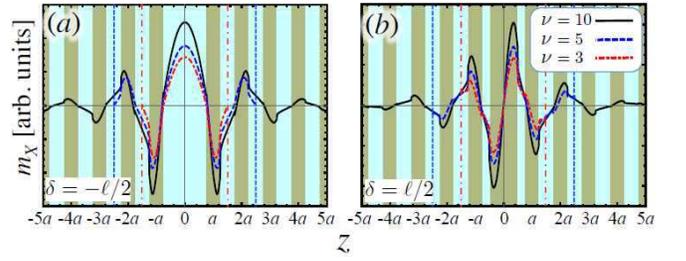}
\caption{Defect mode for different values of $\nu$.
In ($a$) the case $\delta=-\ell/2$ is depicted, where the SW excitation exhibits a symmetric profile around the modified stripes, while in ($b$) an antisymmetric SW profile is observed for $\delta= \ell/2$.
The vertical dashed (dotted-dashed) line depicts the unit cell for $\nu=5$ ($\nu=3$).}
\label{Fig5}
\end{figure}

On the other side, it is possible to see that the defect modes are always revealing a periodic dispersion with finite oscillation amplitude.
Nonetheless, this amplitude decreases dramatically as the lattice parameter of the MSL $\nu a$ increases.
This is depicted in Fig.\ \ref{Fig4}($a$), where the cases $\nu=3$, 5 and 10 are shown.
One can observe that the position of the defect modes is not significantly affected by $\nu$, nevertheless the oscillation amplitude $A_f$ [defined in Fig.\ \ref{Fig4}($a$)] and the number of peaks are clearly dependent on $\nu$.
Thus, at $\nu=10$ for instance the mode inside the BG seems to have no dispersion, which is in agreement with recent BLS experiments and micromagnetic simulations.\cite{Di14a,Zhang16}
Fig.\ \ref{Fig4}($b$) shows a zoom of Fig.\ \ref{Fig4}($a$) around one boundary of the first Brillouin zone ($k_z=\pi/a$), where a finite oscillation amplitude $A_f$ is observed.
The behavior of $A_f$ as a function of $\nu$ is illustrated in Fig.\ \ref{Fig4}($c$), where the oscillation amplitude decreases exponentially as $\nu$ increases.

The spatial spin-wave profiles of the defect modes located within the first band gap, obtained from the in-plane dynamic component $m_X$, are depicted in Fig.\ \ref{Fig5}  for $\delta=\pm \ell/2$ and $\nu=3$, 5 and 10.
The vertical dashed-dotted (dashed) line depicts the unit cell for $\nu=3$ ($\nu=5$).
An important conclusion from Fig.\ \ref{Fig5} is that in addition to the reported antisymmetric defect states in Refs.\ \onlinecite{Di14a} and \onlinecite{Zhang16}, the MSL can be tuned by modifying the width of the defect stripe in such a way that the nature of the defect mode is either symmetric or antisymmetric.
For instance, if $\delta<0$ the mode is symmetric as shown Fig.\ \ref{Fig5}($a$), whereas is antisymmetric when $\delta>0$ [see Fig.\ \ref{Fig5}($b$)].
Note that these symmetry properties are valid for other kind of magnetic materials as long as the defect stripe corresponds to the one with lower saturation magnetization, since if the materials A and B are exchanged, these symmetry properties are also reversed (not shown).
On the other side, unlike the defect modes, where the excitation is mainly located at the defect zone, the branches at the band-gap edges show an extended character, in such a way that these branches are excited in the entire unit supercell (see Ref.\ \onlinecite{Di14a} for details).

From Fig.\ \ref{Fig5} it is easy to see that the SW profile of the defect modes decreases quickly as $z$ increases and this effect is enhanced as $\nu$ increases.
Thus, for $\nu=10$, the SW excitation is almost zero at $z=\pm5a$. 
This localization of the defect mode allows implementing the following: 
If the width of the fifth stripe (localized at $z=\pm5a$) in the lattice with $\nu=10$ is geometrically modified, both the frequency and localization of the defect mode obviously should not change, because the area around $z=\pm5a$ is irrelevant for the dynamics of both S or AS modes. 
To corroborate this behavior, one may employ the case $\nu=10$ with $\delta>0$, in such a way to excite the AS mode, and at the same time modify the width of the fifth stripe by changing $\ell\rightarrow \ell +2\delta^{\prime}$ (with $\delta^{\prime}<0$), in order to excite simultaneously both S and AS modes inside the magnonic band gap.
\begin{figure}[t]
\includegraphics[width=0.48\textwidth]{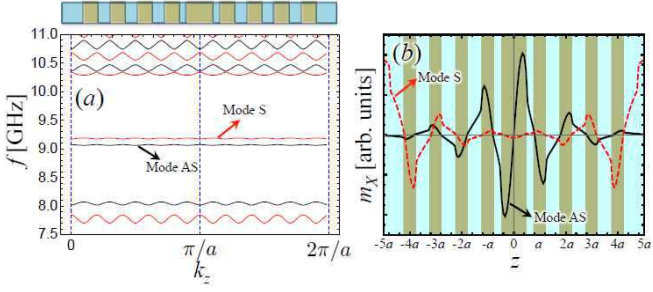}
\caption{In ($a$) a superlattice structure with $\nu=10$, $\delta= \ell/2$ and $\delta^{\prime}=-\ell/2$ is depicted, while ($b$) shows the dynamic magnetization component $m_X$ of both S and AS modes.}
\label{Fig6}
\end{figure}
The calculation of a superlattice with two alternating widths $\ell +2\delta$ and $\ell +2\delta^{\prime}$ for each fifth stripe can be implemented in the theory by replacing the term $\cos(n\pi)\sin\left(n\pi \ell/10 a\right)$ in Eq.\ (\ref{FC}) by $\cos(n\pi)\sin\left[n\pi (\ell +2\delta^{\prime})/10 a\right]$, in such a way that $\delta$ modifies the width of the stripe located at $z=j \times10a$ and $\delta^{\prime}$ modifies the width of the stripe in $z=(j+1/2)10a$, with $j=0,1,2,3...$.
In Fig.\ \ref{Fig6} such a superlattice structure with two different defects characterized by $\delta= \ell/2$ and $\delta^{\prime}=-\ell/2$ is shown.
As mentioned above, both the frequency and localization of S and AS modes are not modified [see Figs.\ \ref{Fig2}($f$) and \ref{Fig2}($j$)].
The interesting feature of this kind of system is that clearly uncoupled symmetric and antisymmetric defect modes may be simultaneously excited, which evolve from the upper and lower boundaries of the band gap respectively, as $\delta$ and $\delta^{\prime}$ are increased in magnitude.

\begin{figure}[t]
\includegraphics[width=0.46\textwidth]{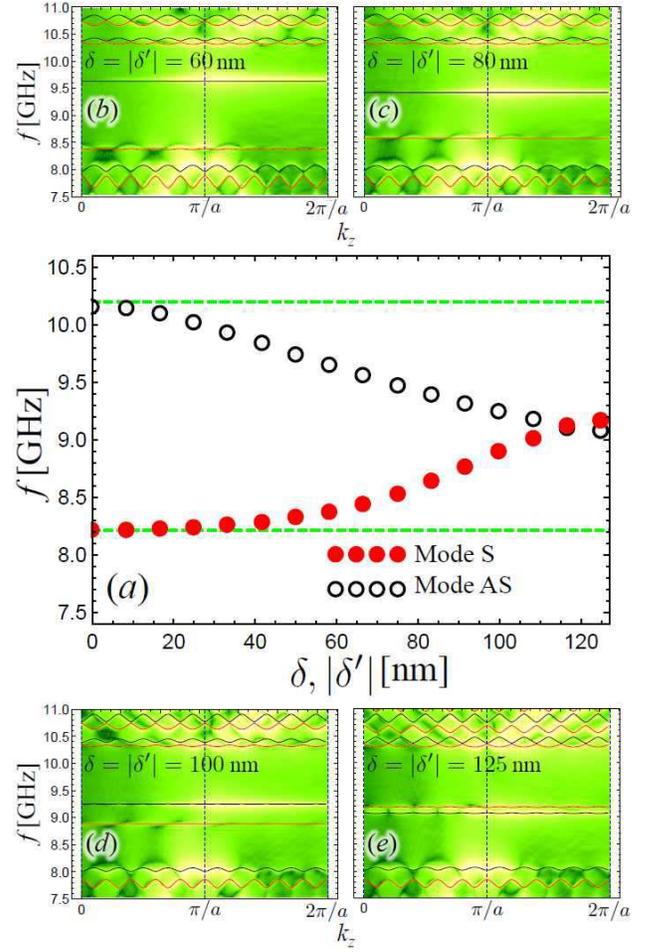}
\caption{a) Symmetric and antisymmetric modes as a function of $\delta$ and $\delta^{\prime}$ (note that for $\delta^{\prime}$ the magnitude is plotted, as $\delta^{\prime}<0$).
Figures (b)-(e) depict the simulated and theoretically calculated SW dispersion for some values of $\delta$ and $|\delta^{\prime}|$.}
\label{Fig7}
\end{figure}


In Fig.\ \ref{Fig7}($a$) the evolution of the S and AS modes for a MSL with two different kind of defects is shown as a function of the magnitude of $\delta$ and $\delta^{\prime}$, where it is assumed that $\delta>0$ and $\delta^{\prime}<0$. 
Figs.\ \ref{Fig7}($b$)--($c$) show the simulated and calculated dispersions for two specific values of $\delta$ and $\vert\delta^{\prime}\vert$, 60 and 80 nm. 
Figs.\ \ref{Fig7}($d$)--($e$) show the simulated and calculated dispersions for $\delta=\vert\delta^{\prime}\vert=100$ and 125 nm. 
Note that there is a crossing point close to $\delta=\vert\delta^{\prime}\vert=115$ nm, where both S and AS modes have the same frequency.
Then, for the case $\delta=\vert\delta^{\prime}\vert=125$ nm, the S mode has a slightly larger frequency than the AS mode, as opposed to the cases where $\delta=\vert\delta^{\prime}\vert<115$ nm. 
One can see in Figs.\ \ref{Fig6} and \ref{Fig7} that two modes appear within the frequency BG.
This is related with the incorporation of two defect stripes in the unit supercell of the superlattice [see Fig.\ \ref{Fig6}($b$)].
Hence, it is expected that an arbitrary distribution of defect stripes in the unit supercell of the superlattice should induce a broad band consisting of multiple nearly flat modes within the band gap.

\begin{figure}[t]
\includegraphics[width=0.48\textwidth]{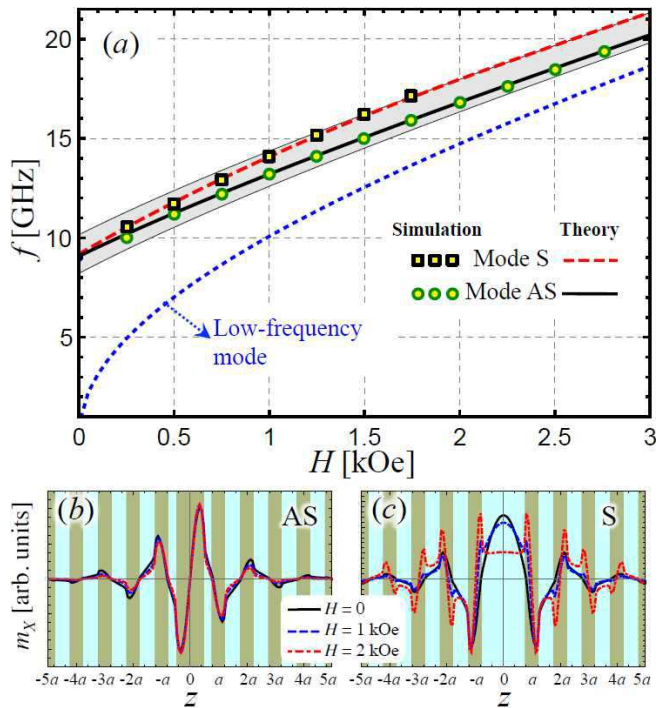}
\caption{ ($a$) Evolution of the symmetric (dashed line) and antisymmetric (solid line) modes as a function of the external field for  $k_z=0$.
Here, the S mode is excited with $\delta= -\ell/2$, and the AS mode with $\delta= \ell/2$.
The gray zone depicts the first band gap and the (blue) dotted line indicate the low-frequency FMR mode.
In ($b$) and ($c$) the SW profiles are depicted for the AS and S mode, respectively, where it is clearly seen that only the S mode in influenced by the field. }
\label{Fig8}
\end{figure}


The proposed MSL with two different defects would be especially interesting for FMR measurements, since the nature of the external excitation in typical FMR setups allows to excite only the symmetric modes, and therefore under specific conditions the S mode should be detected at $k_z=0$. 
The applied field dependence of the S and AS modes is shown in Fig.\ \ref{Fig8}($a$) at the FMR limit ($k_z=0$). 
Here, the low-frequency mode is plotted together with the S and AS modes, where one notices that the symmetric mode is clearly influenced by the field, in such a way that at higher values of $H$ the mode moves towards the high-frequency edge of the band gap (gray zone). 
Nevertheless, the AS mode remains almost in the same relative frequency position with respect to the BG.  
Figs.\ \ref{Fig8}($b$) and \ref{Fig8}($c$) show the SW profiles for the AS and S modes, respectively.
Clearly, the profile of the AS modes remains constant as the external field increases, whereas the S mode is notably modified. 
Since the dynamic part of the Zeeman energy density can be expressed as $\epsilon_{\rm{z}}=-(H/2M_{\rm{s}})(m_X^2+m_Y^2)$, it is expected that the AS modes do not have an additional dynamic contribution in $\epsilon_{\rm{z}}$, since the term $m_X^2+m_Y^2$ is not modified.
Nevertheless, since for the symmetric mode the dynamic component of the magnetization $m_X$ changes with the field, its frequency within the gap is influenced by the external field. 
Therefore, it is demonstrated that the symmetric modes have a limited range of field where they can be observed, since when these modes reach the band-gap edges they are extended along the crystal and therefore they can hardly be detected.\cite{Di14a}
On the other side, once the antisymmetric mode is excited, it should be observable in a wider range of fields.
The same behavior of the defect modes as a function of the field value is valid for different magnitudes of $\delta$ (not shown).

In the theoretical approach it has been assumed that the damping parameter is zero, since this parameter does not significantly affect the band structure (real part of the frequency) of the modes.
Nevertheless, the damping parameter is finite and different in both materials in such a way that the lifetime of the spin waves is dependent of the propagation direction of the waves.\cite{Tiwari10,Vivas12}
Therefore, while the real part of the defect modes is nearly flat, the imaginary part should be dependent on the wave vector.
Nonetheless, the quantitative analysis of the relaxation processes in magnonic superlattices is beyond the scope of this study, since here the main focus are the band structure of magnonic superlattices.


\section{Final remarks}
Dynamic characteristics of one-dimensional bi-component magnonic superlattices have been theoretically studied by taking both the dipolar and the exchange interactions into account.
Symmetry, localization as well as field-dependent properties of the nearly flat defect modes have been theoretically addressed and corroborated by micromagnetic simulations.
It is found that by controlling the width of the modified stripe of the magnonic superlattice either symmetric or antisymmetric modes can be excited.
Also, by modifying the separation between defects, a transition from dispersive to practically flat spin-wave branches is observed inside the magnonic band gaps.
Due to the localization features of the defect modes, a system is proposed that consists of a superlattice with wide and narrow stripe-like defects, where it is possible to observe uncoupled symmetric and antisymmetric modes at the same time.
It is also demonstrated that the symmetric modes have a limited range of fields where they can be observed, while the antisymmetric ones should be externally detected in a wider range of external fields.
The dynamic properties observed in this work can be used to engineer the band structure of magnonic superlattice systems, since the controlled introduction of defects provides additional degrees of freedom, which can be of fundamental importance for technological applications in magnonic crystal based devices. 
\section*{ACKNOWLEDGMENTS}
RAG acknowledges financial support from FONDECYT Iniciacion Grant 11170736 and CONICYT PAI/ACADEMIA Grant 79140033. 
ASN acknowledges funding from FONDECYT Grant 1150072. This work was also supported by FONDECYT Grants 1161403 and 1150072, and the Basal Program for Centers of Excellence,
Grant No. FB0807 CEDENNA, CONICYT.
TS acknowledges funding by the Deutsche Forschungsgemeinschaft (grant GE1202/9-2) and funding from the In-ProTUC scholarship.
ML acknowledge funding by the Deutsche Forschungsgemeinschaft (grant LE2443/5-1). 
Funding from DAAD grant ALECHILE 57136331 and CONICYT PCCI140051 are also highly acknowledged.

\appendix

\section{Effective fields and matrix elements}
\label{ApA}

For the periodic structure shown in Fig.\ \ref{Fig1}, the static exchange field is given by 
\begin{eqnarray}
H^{\rm{ex0}}_{Z}(\mathbf{r})&=&- 4\pi\sum_{\mathbf{G},\mathbf{G}^{\prime}}\mathbf{G}\cdot (\mathbf{G}^{\prime}+\mathbf{G})M_{\rm{s}}(\mathbf{G})[\lambda_{\rm{ex}}(\mathbf{G}^{\prime})]^{2} \times \nonumber
\\
&&e^{i(\mathbf{G}^{\prime}+\mathbf{G})\cdot\mathbf{r}},
\end{eqnarray}
where the other two static components are zero ($H^{\rm{ex0}}_{X}=H^{\rm{ex0}}_{Y}=0$).
On the other side, the dynamic exchange components are
\begin{eqnarray}
h^{\rm{ex}}_{X,Y}(\mathbf{r})&=-&4\pi\sum_{\mathbf{G},\mathbf{G}^{\prime}}(\mathbf{G}+\mathbf{k}) \cdot (\mathbf{G}^{\prime}+\mathbf{G}+\mathbf{k})[\lambda_{\rm{ex}}(\mathbf{G}^{\prime})]^{2}\times \nonumber
\\
&& m_{X,Y}(\mathbf{G})e^{i(\mathbf{G}^{\prime}+\mathbf{G}+\mathbf{k})\cdot\mathbf{r}}.
\end{eqnarray}

According to Fig.\ \ref{Fig1} the external applied field is $H^{0}_{Z}= H\cos (\varphi_{\rm{ h}}-\varphi)$ and $H^{0}_{X}= H\sin (\varphi_{\rm{ h}}-\varphi)$, where $\varphi_{\rm{ h}}$ ($\varphi$) is the angle between the external field (equilibrium magnetization) and the $z$-axis. 
On the other hand, the dynamic components of the dipolar field are
\begin{equation}
h^{\rm{d}}_{Y}(\mathbf{r})=-\sum_{\mathbf{G}}m_{Y}(\mathbf{G})\zeta(\mathbf{G},\mathbf{k})e^{i(\mathbf{G+k})\cdot \mathbf{r}}
\label{hxx}
\end{equation}
and
\begin{equation}
h_{X}^{\rm{d}}(\mathbf{r})=4\pi\sum_{\mathbf{G}}m_{X}(\mathbf{G})\xi(\mathbf{G})^2[\frac{\zeta(\mathbf{G},\mathbf{k})-1}{\vert \mathbf{G}+\mathbf{k} \vert ^2 }]
e^{i(\mathbf{G+k})\cdot \mathbf{r}}.
\label{hx}
\end{equation}
Here, it has been defined $\xi(\mathbf{G})=(G_{n}^{\nu}+k_z)\sin\psi$ and
\begin{equation}
\zeta(\mathbf{G},\mathbf{k})=\frac{2\sinh[\vert \mathbf{G}+\mathbf{k} \vert d/2]e^{-\left\vert  \mathbf{G}+\mathbf{k}  \right\vert d/2}}{\vert  \mathbf{G}+\mathbf{k} \vert d}.
\end{equation}
Also, the $Z$-component of the static dipolar field is
\begin{equation}
H^{\rm{d0}}_{Z}(\mathbf{r})=-4\pi \sum_{\mathbf{G}}M_{\rm{s}}(\mathbf{G})\chi(\mathbf{G})^2
\frac{1-\zeta(\mathbf{G},0)}{\left\vert \mathbf{G}\right\vert^2}e^{i\mathbf{G}\cdot \mathbf{r}},
\label{hy}
\end{equation}
where $\chi(\mathbf{G})=G_{n}^{\nu}\cos \varphi$.

By introducing the effective fields in the dynamic equation of motion, the submatrices in Eq.\ (\ref{subM}) are given by
\begin{widetext}
\begin{subequations}
\begin{eqnarray}
 \mathbf{A}^{XX}_{\mathbf{G},\mathbf{G}^{\prime}}  &=&\mathbf{A}^{YY}_{\mathbf{G},\mathbf{G}^{\prime}}=0,
\\
 \mathbf{A}^{XY}_{\mathbf{G},\mathbf{G}^{\prime}}  &=&-H\cos(\varphi-\varphi_{\rm{h}})\delta_{\mathbf{G},\mathbf{G}^{\prime}}+4\pi M_{\rm{s}}(\mathbf{G}-\mathbf{G}^{\prime})\left[\chi(\mathbf{G}-\mathbf{G}^{\prime})^2 \frac{1-\zeta(\mathbf{G}-\mathbf{G}^{\prime},0)}{\left\vert \mathbf{G}-\mathbf{G}^{\prime} \right\vert^2}-\zeta(\mathbf{G}^{\prime},\mathbf{k})\right]\nonumber
 \\
 &-&4\pi \sum_{\mathbf{G}^{\prime\prime}}M_{\rm{s}}(\mathbf{G}-\mathbf{G}^{\prime\prime})\left[\left(\mathbf{G}^{\prime}+\mathbf{k}\right)\cdot\left(\mathbf{G}^{\prime\prime}+\mathbf{k}\right)-\left(\mathbf{G}-\mathbf{G}^{\prime\prime}\right)\cdot\left(\mathbf{G}-\mathbf{G}^{\prime}\right)\right][\lambda_{\rm{ex}}(\mathbf{G}^{\prime\prime}-\mathbf{G}^{\prime})]^{2} 
 \\
 \mathbf{A}^{YX}_{\mathbf{G},\mathbf{G}^{\prime}}   &=&
 H\cos(\varphi-\varphi_{\rm{ h}})\delta_{\mathbf{G},\mathbf{G}^{\prime}}
 -4\pi M_{\rm{s}}(\mathbf{G}-\mathbf{G}^{\prime})\Bigg[\chi(\mathbf{G}-\mathbf{G}^{\prime})^2\frac{1-\zeta(\mathbf{G}-\mathbf{G}^{\prime},0)}{\left\vert \mathbf{G}-\mathbf{G}^{\prime} \right\vert^2}
 +\xi(\mathbf{G}^{\prime})^2[\frac{\zeta(\mathbf{G}^{\prime},\mathbf{k})-1}{\vert \mathbf{G}^{\prime}+\mathbf{k} \vert ^2 }]\Bigg]\nonumber
 \\
 &+&4\pi \sum_{\mathbf{G}^{\prime\prime}}M_{\rm{s}}(\mathbf{G}-\mathbf{G}^{\prime\prime})\left[\left(\mathbf{G}^{\prime}+\mathbf{k}\right)\cdot\left(\mathbf{G}^{\prime\prime}+\mathbf{k}\right)-\left(\mathbf{G}-\mathbf{G}^{\prime\prime}\right)\cdot\left(\mathbf{G}-\mathbf{G}^{\prime}\right)\right][\lambda_{\rm{ex}}(\mathbf{G}^{\prime\prime}-\mathbf{G}^{\prime})]^{2}. \end{eqnarray}
\label{AA}
\end{subequations}
\end{widetext}

\section{Fourier coefficient of one-dimensional magnonic superlattices}
\label{ApB}

For a general one-dimensional superlattice, the Fourier coefficient of the saturation magnetization can be obtained by analyzing the one-dimensional periodic structure.
Thus, according to Fig.\ \ref{Fig1}, it is straightforward to see that
\begin{widetext}
\begin{eqnarray}
M_{\rm{s}}(G_{n}^{\nu})&=&\frac{1}{2 \nu a}
\left[M^{\rm{A}}_{\rm{s}}\int_{-\frac{\nu a}{2}}^{-\frac{\nu a-\ell}{2}}e^{-iG_{n}^{\nu} z}dz\right. 
+M^{\rm{B}}_{\rm{s}}\int_{-\frac{\nu a-\ell}{2}}^{-\frac{(\nu-2)a+ \ell}{2}}e^{-iG_{n}^{\nu} z}dz
+M^{\rm{A}}_{\rm{s}}\int_{-\frac{(\nu-2)a+ \ell}{2}}^{-\frac{(\nu-2)a-\ell}{2}}e^{-iG_{n}^{\nu} z}dz \nonumber
\\
&+&...+M^{\rm{A}}_{\rm{s}}\int_{-\frac{\ell +2\delta}{2}}^{\frac{\ell+2\delta}{2}}e^{-iG_{n}^{\nu} z}dz+...+
M^{\rm{B}}_{\rm{s}}\int_{\frac{(\nu-2)a+ \ell}{2}}^{\frac{\nu a-\ell}{2}}e^{-iG_{n}^{\nu} z}dz
+\left.M^{\rm{A}}_{\rm{s}}\int_{\frac{\nu a-\ell}{2}}^{\frac{\nu a}{2}}e^{-iG_{n}^{\nu} z}dz\right],
\label{A1}
\end{eqnarray}
\end{widetext}
if $\nu$ is an even number. 
On the other side, if $\nu$ is an odd number, the coefficient is calculated as
\begin{widetext}
\begin{eqnarray}
M_{\rm{s}}(G_{n}^{\nu})&=&\frac{1}{2 \nu a}
\left[M^{\rm{B}}_{\rm{s}}\int_{-\frac{\nu a}{2}}^{-\frac{(\nu-1) a+ \ell}{2}}e^{-iG_{n}^{\nu} z}dz\right. \nonumber
+M^{\rm{A}}_{\rm{s}}\int_{-\frac{(\nu-1) a+ \ell}{2}}^{-\frac{(\nu-1) a-\ell}{2}}e^{-iG_{n}^{\nu} z}dz\nonumber
+M^{\rm{B}}_{\rm{s}}\int_{-\frac{(\nu-1) a-\ell}{2}}^{-\frac{(\nu-3) a+ \ell}{2}}e^{-iG_{n}^{\nu} z}dz \nonumber
\\
&+&...+M^{\rm{A}}_{\rm{s}}\int_{-\frac{\ell +2\delta}{2}}^{\frac{\ell +2\delta}{2}}e^{-iG_{n}^{\nu} z}dz+...+
M^{\rm{A}}_{\rm{s}}\int^{\frac{(\nu-1) a+ \ell}{2}}_{\frac{(\nu-1) a-\ell}{2}}e^{-iG_{n}^{\nu} z}dz
+\left.M^{\rm{B}}_{\rm{s}}\int_{\frac{(\nu-1)a+ \ell}{2}}^{\frac{\nu a}{2}}e^{-iG_{n}^{\nu} z}dz\right] .
\label{A2}
\end{eqnarray}
\end{widetext}

Then, by carrying out the appropriate integration of Eqs. (\ref{A1}) and (\ref{A2}), the result can be readily  generalized as
\begin{widetext}
\begin{equation}
M_{\rm{s}}(G_{n}^{\nu})=M_{\rm{s}}^{\rm{B}}\frac{\sin(n\pi)}{n\pi}+\frac{M_{\rm{s}}^{\rm{A}}-M_{\rm{s}}^{\rm{B}}}{n\pi}\left\{\sin\left[\frac{n\pi(\ell +2\delta)}{\nu a}\right]+\Psi(n,\nu)\sin\left(\frac{n\pi \ell}{\nu a}\right)\right\} ,
\label{FC}
\end{equation}
\end{widetext}
where
\begin{widetext}
\begin{eqnarray}
\Psi(n,\nu)&=&\cos(n\pi)\cos^2\left(\nu\frac{\pi}{2}\right)-2+2\sum_{j=1}^{\nu}\left\{\cos\left[\frac{(j-1)n\pi}{\nu}\right]\cos^2\left[(j+1)\frac{\pi}{2}\right]\cos^2\left[(\nu+1)\frac{\pi}{2}\right]\right. \nonumber
\\
&+&
\left.\cos\left[\frac{(j-2)n\pi}{\nu}\right]\cos^2\left(j\frac{\pi}{2}\right)\cos^2\left(\nu\frac{\pi}{2}\right)\right\}.
\label{MS}
\end{eqnarray}
\end{widetext}
Here, $\nu$ represents the number of lattice repetitions that are necessary to form the MSL.
A similar structure can be used for the exchange length $\lambda_{\rm{ex}}(G_{n}^{\nu})$.
Therefore, by choosing $\delta$ and $\nu$ any one-dimensional bi-component magnonic superlattice can be modeled.


%

\end{document}